# Topological Electronic Structure and Transport Properties of the Distorted Rutile-type WO$_2$

Yuto Muramatsu[1], Daigorou Hirai[1], Mitsuaki Kawamura[2], Susumu Minami[3], Yoshitaka Ikeda[3],

Takahiro Shimada[3], Keita Kojima[4], Naoyuki Katayama[1], and Koshi Takenaka[1]

[1]*Department of Applied Physics, Nagoya University, Nagoya 464–8603, Japan*
[2] *Information Technology Center, University of Tokyo, Bunkyo-ku, Tokyo 113-8658, Japan*
[3] *Department of Mechanical Engineering and Science, Kyoto University, Nishikyo-ku, Kyoto 615-8540, Japan*
[4] *Institute for Solid State Physics, the University of Tokyo, Kashiwa 277-8581, Japan*

Abstract: We elucidate the transport properties and electronic structures of distorted rutile-type WO$_2$. Electrical resistivity and Hall effect measurements of high-quality single crystals revealed the transport property characteristics of topological materials; these characteristics included an extremely large magnetoresistance of 13,200% (2 K and 9 T) and a very high carrier mobility of 25,700 cm$^2$ V$^{-1}$ s$^{-1}$ (5 K). First-principles calculations revealed Dirac nodal lines (DNL) near the Fermi energy in the electronic structure when spin-orbit interactions (SOIs) were absent. Although these DNLs mostly disappeared in the presence of SOIs, band crossings at high-symmetry points in the reciprocal space existed as Dirac points. Furthermore, DNLs protected by nonsymmorphic symmetry persisted on the $k_y = \pi/b$ plane. The unique transport properties originating from the topological electronic structure of chemically and thermally stable WO$_2$ could represent an opportunity to investigate the potential electronic applications of the material.

## I. INTRODUCTION

In recent years, topological semimetals have attracted considerable attention in both basic science and practical applications. Owing to their unique band topology, topological semimetals exhibit linear band dispersion and unusual transport properties, such as large magnetoresistance (MR) and ultrahigh mobility [1] [2] [3] [4] [5]. As a product of these characteristics, these materials are promising candidates for next-generation electronic applications.

For industrial applications, of the absence of toxic elements and thermal and chemical stabilities are highly desirable. Therefore, thermally and chemically stable transition-metal oxides are promising candidates. For industrial applications, ideal candidate materials exhibit thermal and chemical stability and are devoid of toxic elements; promising candidates therefore include thermally and chemically stable transition metal oxides. For instance, RuO$_2$ and IrO$_2$ with rutile-type structures in the *P*4$_2$/*mnm* space group have two types of Dirac nodal lines (DNLs) [6] [7]; although experimentally observed, one is protected, and the other is unprotected by nonsymmorphic symmetry [8]. An energy gap opens in unprotected DNLs owing to spin-orbit interactions (SOIs). A theoretical study suggested that this energy gap in the DNLs induces the large spin Hall effect (SHE) observed in the amorphous and thin films of IrO$_2$ [8] [9] [10].In another example, *β*-ReO$_2$ with a *α*-PbO$_2$ structure has the "Dirac Loop Chain," a unique topological electronic structure, that is a chain of orthogonal DNLs connected via points [11]. The large MR of 22,000% observed at 2 K and 10 T was attributed to this topological electronic structure [12] [13].

MoO$_2$ and WO$_2$ have rutile-related crystal structures; dimer formation between the transition metals results in a distorted-rutile-type structure with a *P*2$_1$/*c* space group, as shown in Fig. 1(a) [14] [15] [16]. Whether MoO$_2$ and WO$_2$ have DNLs similar to those of rutile compounds, such as RuO$_2$ and IrO$_2$ is unclear because the symmetry that protects the DNLs is lost owing to lattice distortion. Recently, MoO$_2$ was reported to exhibit a large MR (5.3 × 10$^4$% at 2 K and 9 T) [17]. In addition, a large SHE comparable to that of IrO$_2$ was observed for the thin films of WO$_2$ [18]. These experimental results suggest the existence of DNLs in distorted-rutile compounds despite the presence of SOIs as strong as 0.5 eV in the 5*d* electron system [19]. The elucidation of the existence of DNLs and the mechanism of DNL formation in WO$_2$, in terms of the development of electronic and spintronic materials, constitutes a matter of significant interest.

In this study, resistivity, MR, and Hall effect measurements were performed on high-quality WO$_2$ single crystals grown using the chemical vapor transport method. The MR increased significantly with an increasing magnetic field, and a large MR of 13,200% was observed at 2 K and 9 T. Hall effect measurements demonstrated carrier compensation and the presence of hole carriers with considerably high mobility of 25,700 cm$^2$ V$^{-1}$ s$^{-1}$. These results indicate the presence of a linear dispersion near the Fermi energy ($E_F$). *Ab-initio* calculations revealed the presence of Dirac points and DNLs protected by the helical axis against strong SOIs.

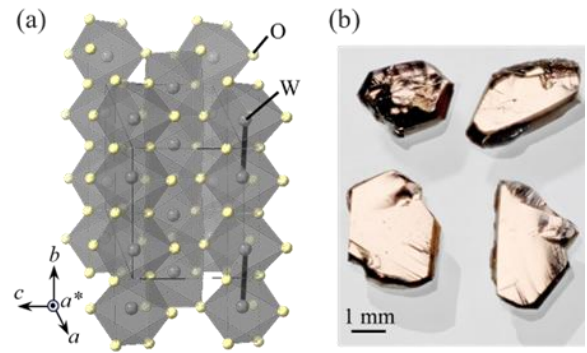

Fig. 1. (a) Crystal structure of WO$_2$, which is a rutile structure with monoclinic distortion due to the dimerization of W atoms (black solid lines). (b) Photograph of single crystals of WO$_2$ grown by using a chemical vapor transport method.

## II. EXPERIMENTAL METHOD

Single crystals of WO$_2$ were grown by using a chemical vapor transport method. WO$_2$ powder was sealed in a quartz tube with iodine as a transport agent and placed in a three-zone furnace with a temperature gradient of 750–850 K for five days. The obtained single crystals had a typical size of 1.5 × 2.0 × 0.5 mm$^3$ [Fig. 1(b)]. Many large crystals were twinned. The size of the single-domain crystal used for measurements was 1.0 × 0.5 × 0.3 mm$^3$. The powdered single crystals of WO$_2$ were utilized for characterization by performed powder X-ray diffraction (XRD) measurements with Cu Kα radiation using a diffractometer (MiniFlex, Rigaku). The lattice constants and the orientation of a single crystal were determined by performing single-crystal XRD measurements using Mo-Kα radiation on a diffractometer (XtaLAB mini II, Rigaku). The long axis of the rod-shaped crystals corresponded to the $a^*$ axis, whereas the $b$ and $c$ axes were perpendicular to the $a^*$ axis.

Electrical resistivity and Hall effect measurements were performed using a physical property measurement system (PPMS, Quantum Design). The electrical resistivity was measured using a four-probe method with an electrical current applied in the $a^*$ direction and magnetic fields applied in the $b$ or $c$ direction. The angular dependency of the longitudinal electrical resistivity was evaluated using a sample rotator equipped with a PPMS. The Hall effect was examined using the five-probe method, and $\rho_{yx}(B)$ was obtained as $\rho_{yx}(B) = [\rho_{yx}(+B) − \rho_{yx}(−B)]/2$ to ignore the longitudinal contribution due to misalignment of the contacts.

The electronic structure of WO$_2$ was analyzed using the Vienna Ab initio Simulation Package code [20,21]. The lattice constants and atomic coordinates used for the initial calculations were determined from synchrotron XRD measurements at 104 K (BL02B1 of SPring-8). The obtained lattice constants were $a$ = 5.55715(2) Å, $b$ = 4.89890(2) Å, $c$ = 5.57190(2) Å, $α$ = 90°, $β$ = 118.8574(6)°, and $γ$ = 90°, which are consistent with those reported previously [14]. The corresponding crystallographic information (CIF) file is included in the Supplementary Material. We used a plane-wave basis with a cut-off of 520 eV to represent the Kohn–Sham orbitals. The Perdew–Burke–Ernzerhof generalized-gradient approximation [22] was employed for the exchange-correlation functional. A 12 × 16 × 12 $k$-point mesh with the Monkhorst–Pack scheme [23] was used for Brillouin zone (BZ) sampling. The optimized lattice constant of WO$_2$ was $a$ = 5.59 Å, $b$ = 4.49 Å, and $c$ = 5.60 Å, consistent with the experimental results. Based on the obtained Bloch states, we constructed a Wannier function to analyze the nodal lines and Fermi surface. A Wannier basis set was constructed using the Wannier90 code [24]. The basis consists of $d$-character orbitals localized at the W site, that is, 20 (40) orbitals without (with) the spin-orbit coupling (SOC). The Fermi surfaces were displayed using FermiSurfer [25].

## III. RESULTS

### 3.1 Extremely large magnetoresistance

Figure 2(a) shows the temperature dependence of electrical resistivity in various magnetic fields. $\rho_{xx}(T)$ in zero magnetic field exhibits a metallic behavior with a considerably small residual resistivity of $\rho_{xx}(2\ K)$ = 236 nΩ cm. This is comparable to $\rho_{xx}(2\ K)$ = 206 nΩ cm of a Dirac loop chain metal $β$-ReO$_2$ [12] and $\rho_{xx}(4.2\ K)$ = 291 nΩ cm of high-quality Bi single crystals [26]. High conductivity at low temperatures suggests the presence of high-mobility carriers. The residual resistivity ratio (RRR), defined as $\rho_{xx}(300\ K)/\rho_{xx}(2\ K)$, is 362, which is more than one order higher than the previously reported value of 15 [14]. The significantly improved RRR signifies the high quality of the WO$_2$ single crystals in this study.

When a transverse magnetic field is applied perpendicular to the current direction ($B \parallel c$, $I \parallel a^*$), the resistivity reaches a minimum, increases with decreasing temperature, and saturates at low temperatures. The minimum temperature increased with the applied magnetic field and reached approximately 80 K at 9 T. Similar behavior was observed in various topological semimetals, including WTe$_2$ [4], NbP [2], and LaSb [27].

The inset in Fig. 2(a) depicts the magnetic field ($B$) dependency of MR defined as $[\rho_{xx}(B) − \rho_{xx}(0)]/\rho_{xx}(0)$. The MR increased rapidly with an increasing magnetic field and reached a high positive value of 13,200% at 2 K and 9 T. The data were fitted by applying a power law to determine that the MR was dependent on $B^{1.6}$. Such a large MR was observed in both topologically nontrivial and trivial semimetals [1] [4] [28] [29] [30] [31] [32]. In Dirac fermions, the MR exhibits a linear field dependence [33], whereas in compensated semimetals, the MR depends on the square of the magnetic field [4] [34] [35]. The MR of WO$_2$ was different from those of both models. In addition, the MR did not saturate at < 9 T. This behavior differs from that of compensated semimetals, in which the MR saturates at high magnetic fields [34] [35].

To understand the carrier-scattering mechanism, the MR data were analyzed using a Kohler plot, as shown in Fig. 2(b). When the carriers follow a conduction mechanism described by a single relaxation time, $τ$, the MR follows Kohler's rule: MR = $a(B/\rho_{xx}(0))$ [36] [37]. The MR data at 50, 75, and 100 K merged into one universal curve [Fig. 2(b), inset], whereas the MR data at 5 and 25 K deviated from the curve. Thus, electrical conduction at above 50 K is described by a single relaxation time, whereas conduction at low temperatures cannot be described by the single relaxation time. The change in the scattering mechanism is also indicated by the change in the magnetic-field angular dependency of the MR. Fig. 2 (c) shows the angular dependency of the resistivity at various temperatures from 2 to 120 K in a polar plot. A current was applied parallel to the $a^*$ axis ($I \parallel a^*$) under a magnetic field of $B$ = 10 T in the $bc$ plane. At 120 and 100 K, the resistivity was almost angle-independent. In contrast, at 80 K, a slight deformation from a circle was observed, and at temperatures below 40 K, a clear anisotropy appeared. The angular dependency of the resistivity at various temperatures of < 40 K shows similar behavior, with maxima at approximately 45° and 135° and minima at 0° and 90°. The angular dependence is close to 4-fold symmetry, reflecting the rutile crystal structure. More precisely, it has 2-fold symmetry due to the monoclinic distortion of the crystal structure. The Kohler plot and angular dependency of the resistivity suggest that at low temperatures, where the MR is more pronounced, multiple types of carriers with different mobilities (relaxation times) are involved in the transport properties.

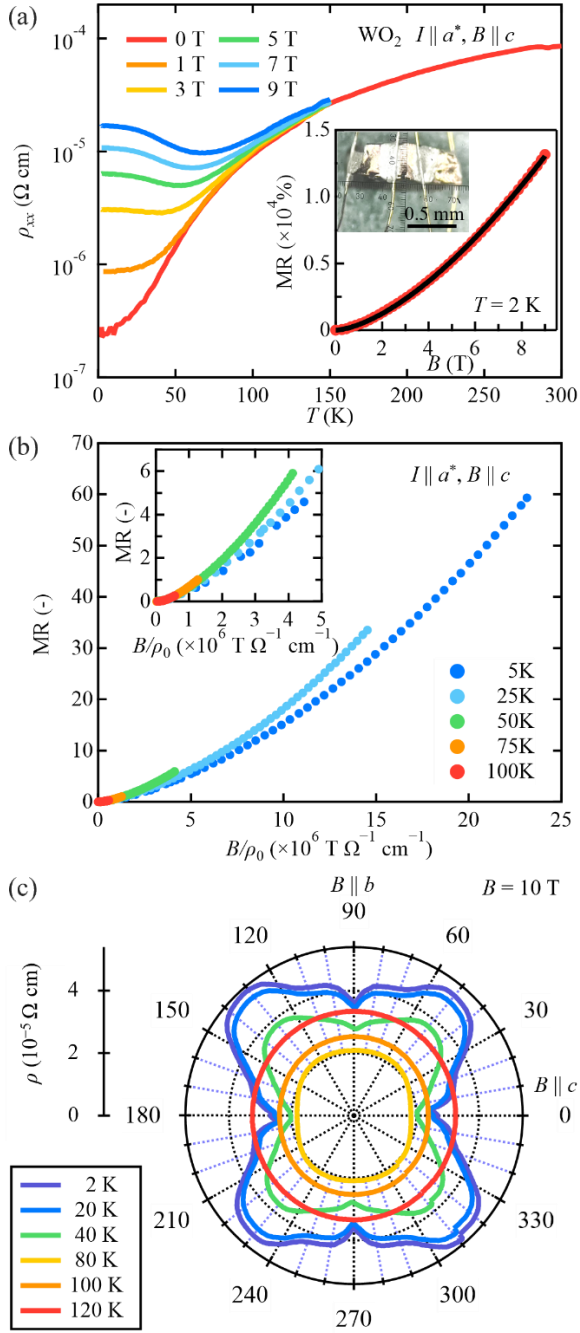

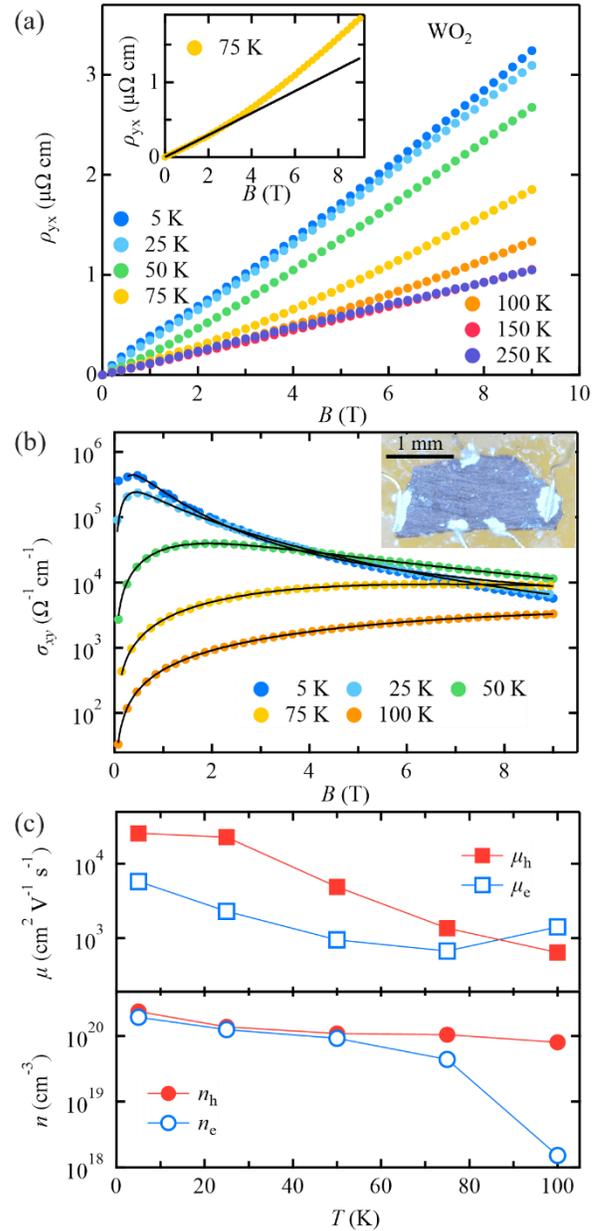

Fig. 2. (a) Temperature dependence of resistivity $\rho_{xx}$ measured with the electrical current flowing along the $a^*$ axis at various magnetic fields in the $c$ axis. The inset shows a photograph of a WO$_2$ sample with electrical contacts for $\rho_{xx}$ measurements and magnetic field dependency of magnetoresistance (MR) measured at 2 K. The black solid line is fit to the power law with an exponent of 1.6. (b) Kohler's scaling plot at various temperatures. The inset shows the violation of the Kohler's rule at < 25 K. (c) Angular plot of resistivity in an applied magnetic field of 10 T measured at various temperatures between 2 and 120 K. 0° and 90° are defined as the magnetic field parallel to the $c$ and $b$ axes ($B \parallel c$ and $b$), respectively.

Fig. 3. (a) Field dependencies of Hall resistivity $\rho_{yx}(B)$ for WO$_2$ measured at various temperatures from 5 to 250 K. The inset shows $\rho_{yx}(B)$ at 75 K deviating from the straight line at high fields. (b) Components of the conductivity tensor, $\sigma_{xy}$, as functions of the magnetic field for temperatures ranging from 5 to 100 K. Black solid lines are the fitting curves based on the two-carrier model. The inset shows a photograph of a WO$_2$ sample with electrical contacts for Hall effect measurements. (c) Temperature dependence of electron and hole carrier mobility ($\mu_e$ and $\mu_h$) and carrier density ($n_e$ and $n_h$) obtained from the two-carrier model fitting shown in (b).

## 3.2 Carrier Compensation and High-Mobility Carriers

To clarify the origin of the large MR, the carrier density and mobility were estimated by performing Hall-effect measurements. At all temperatures, $\rho_{yx}$ is positive and increases with an increasing magnetic field ($B$). $\rho_{yx}$ values at 200 and 250 K are linear against $B$ and nonlinear at temperatures below 100 K in the high-field region [Fig. 3(a) inset]. This indicated the coexistence of electrons and holes. Therefore, the carrier density and mobility were analyzed using a two-carrier model. Fig. 3(b) shows the Hall conductivity obtained from the transverse MR and the fitting curves assuming the following two-carrier model:

$\sigma_{xy} = [n_h\mu_h^2/(1+\mu_h^2B^2) - n_e\mu_e^2/(1+\mu_e^2B^2)]eB$, where $n_e$ and $n_h$ are electron and hole densities, respectively, and $\mu_e$ and $\mu_h$ are electron and hole mobilities, respectively. The obtained temperature dependency of the electron density, hole density, electron mobility, and hole mobility are shown in Fig 3(c). The electron density increases significantly at < 100 K and reaches $1.9 \times 10^{20}$ cm$^{-3}$ at 5 K. The hole density shows a gradual increase and reaches $2.3\times10^{20}$ cm$^{-3}$ at 5 K, which is approximately equivalent to $n_e$. Thus, the hole and electron carriers are almost compensated at low temperatures. The mobility also increases significantly at low temperatures for both electrons and holes, with high values of $\mu_e = 5.7\times10^3$ cm$^2$ V$^{-1}$ s$^{-1}$ and $\mu_h = 2.6\times10^4$ cm$^2$ V$^{-1}$ s$^{-1}$ at 5 K. The $\mu_h$ is as high as the extremely high mobilities of Weyl semimetals MoTe$_2$ ($5.8\times10^4$ cm$^2$ V$^{-1}$ s$^{-1}$) [38] and WTe$_2$ ($1.1\times10^4$ cm$^2$ V$^{-1}$ s$^{-1}$) [39]. The mobility of holes is always greater than that of electrons at low temperatures, resulting in a dominant contribution of holes to conduction. This result is consistent with the observed positive Hall coefficient.

3.3 Electronic structure calculations

First-principles calculations were performed to understand the transport properties of WO$_2$. Figs. 4(a) and (b) show the band dispersions near the Fermi energy of WO$_2$ with and without SOC, respectively. The dominant contribution to the electronic structure near $E_F$ originates from the $t_{2g}$ manifold of the 5$d$ orbitals in the W$^{4+}$ ions. No significant difference exists in the electronic structure from −3 to 2 eV between with and without SOC. However, some of the degenerated bands cross the $E_F$ split by the SOC. The bands with indices 1, 2, and 3 cross the $E_F$, with band 3 forming a large electron-like Fermi surface centered at the $\Gamma$ point and bands 1 and 2 forming hole-like Fermi surfaces centered at the $B$ point [Fig. 4(c)]. The coexistence of electron- and hole-like Fermi surfaces is consistent with the carrier compensation revealed by Hall-effect measurements. Furthermore, the carrier concentrations estimated from the calculated Fermi surface are consistent with the experimentally obtained carrier concentrations for electrons and holes.

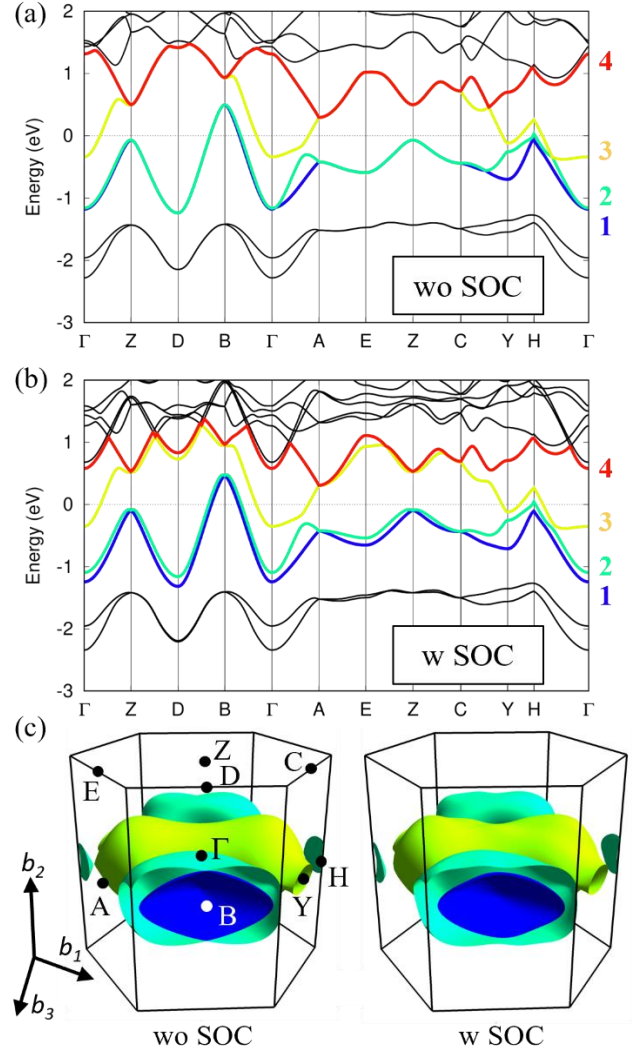

Fig. 4 Band structure of WO$_2$ evaluated (a) without and (b) with SOC. Band indices from 1 to 4 for the bands near $E_F$ are also shown. In (a) and (b), the energy 0 is set to $E_F$. (c) Fermi surface of WO$_2$ determined without and with SOC. The color of the Fermi surface corresponds to that of band dispersion shown in the band structure in (a) and (b).

When the SOC is not considered, the bands with indices 1 and 2 degenerate in many parts of the reciprocal lattice space such as the $Z$-$D$-$B$ and $A$-$E$-$Z$ paths, owing to the symmetry requirements [Fig. 4(a)]. This band degeneracy creates intersections between different bands, forming Dirac points and DNLs. Fig. 5(a) and (b) present the top view of the Fermi surface and its cross-section in the $k_y = 0$ plane consisting of bands 1, 2, and 3. Several intersections exist between bands 1 and 2 and 2 and 3 indicated by red dots; at these points DNLs cross $E_F$. Three-dimensional plots of the energy-dispersive DNLs formed by the intersections between bands 1 and 2 and 2 and 3 are shown in Figs. 5(c) and (d). The points where the two DNLs cross $E_F$ in Figs. 5(c) and (d) correspond to the band intersections depicted in Fig. 5(b).

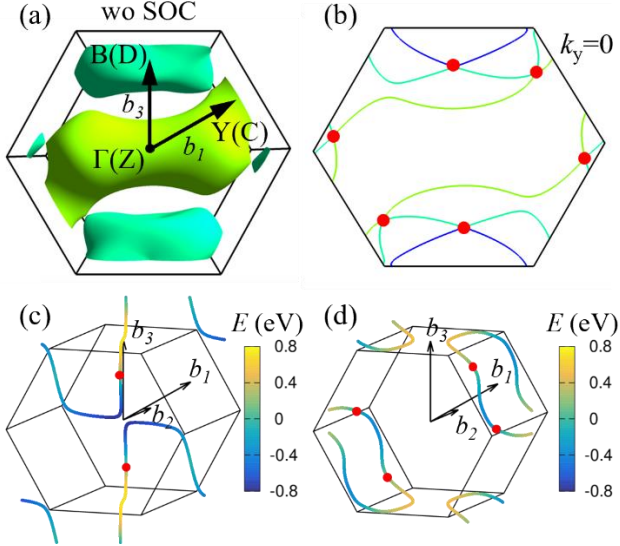

Fig. 5 (a) Fermi surface of WO$_2$ calculated without SOC viewed from the $k_y$ direction. (b) The cross section of Fermi surface of WO$_2$ at $k_y = 0$. Bands 1 and 2, and 3 and 3 intersect each other (red dots) where energy dispersive DNLs cross $E_F$. (c)(d) The energy dispersive DNLs in the electronic structure of WO$_2$ without SOC.

When SOC is considered, the degeneracy of the bands is lifted and the DNLs are significantly affected, as shown in Fig. 4(b); bands 1 and 2, which are degenerated without SOC [Fig. 4(a)], split over most of the $k$-space. However, at high-symmetry points $B$ and $A$, the degeneracies remain protected. The DNLs displayed in Figs. 5(c) and (d) largely disappear as the degeneracies of bands 1, 2, and 3 are lifted, and the gaps open at the band intersections. However, the degeneracy at $B$ and $A$ is maintained; the band crossing remains at the point where the DNLs without SOC cross the boundary of BZ, which forms Dirac points near $E_F$, as shown in Figs. 6(a) and (b). In addition to these Dirac points, DNLs that are robust against SOC exist on the $k_y = \pi/b$ plane. Fig. 6(c) reveals the DNLs on the $k_y = \pi/b$ plane, constructed by bands 1 and 2. These DNLs are protected by the nonsymmorphic symmetry of the distorted rutile-type structure ($2_1$ screw axis parallel to the $b$ axis). Such DNLs protected by $2_1$ screw axes have also been reported for the CaSb$_2$ DNL metal [40].

## IV. DISCUSSION

In WO$_2$, which crystallizes in a distorted rutile-type structure, the DNLs protected by the nonsymmorphic symmetry of the rutile structure, as in IrO$_2$ and RuO$_2$, are lost. Most of the DNLs present in WO$_2$ without SOC are lost because of the gap opened by SOC. However, *ab-initio* calculations show that degeneracy is protected at high-symmetry points in the $k$-space, and some of the DNLs remain as Dirac points. Furthermore, a DNL protected by a nonsymmorphic symmetry ($2_1$ screw axis) exists on the $k_y = \pi/b$ plane. This topological electronic structure near $E_F$ is likely responsible for the unique transport properties of WO$_2$.

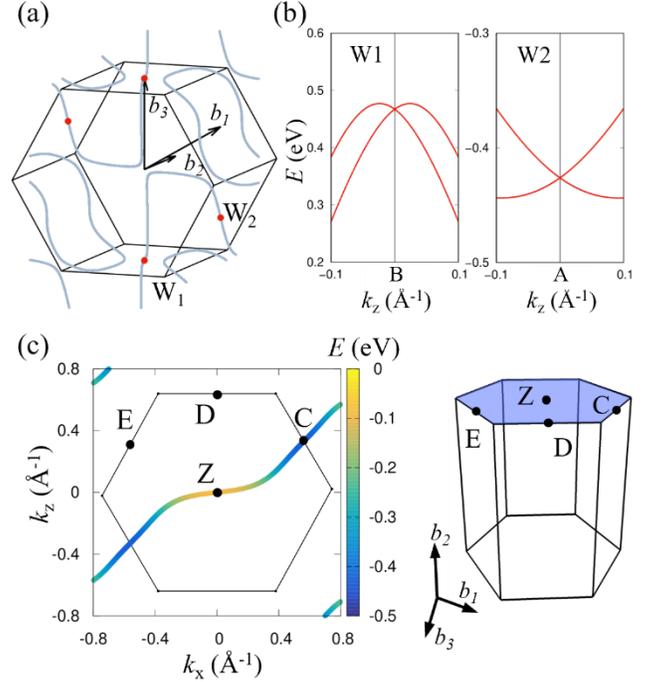

Fig. 6 (a) Preserved Dirac points (red dots) in the presence of SOC. Light-gray lines are DNLs that disappear in the presence of SOC. (b) Energy band dispersion along the $k_z$ direction near the Dirac points. (c) DNLs on the $k_y = \pi/b$ plane protected by nonsymmorphic symmetry against SOC. Bands 1 and 2 in Fig. 4 make DNLs.

We compare the transport properties of WO$_2$ with those of representative topological metals (Table I). We observed an extremely large MR in WO$_2$, which is commonly observed in topological semimetals, such as Dirac and Weyl semimetals. The large MR observed in WO$_2$ is an order of magnitude smaller than those of typical topological semimetals Cd$_3$As$_2$, NbP, and ZrSiS and comparable with that of $\beta$-ReO$_2$. Another characteristic of topological semimetals is ultrahigh mobility of carriers, such as $10^6$ cm$^2$ V$^{-1}$ s$^{-1}$ reported for Cd$_3$As$_2$ and NbP. The mobility of carriers in WO$_2$ at 5 K reaches $2.6 \times 10^4$ cm$^2$ V$^{-1}$ s$^{-1}$, which is one or two orders of magnitude smaller than those of Cd$_3$As$_2$ and NbP, comparable to that of ZrSiS, and one order of magnitude larger than that of $\beta$-ReO$_2$.

The smaller MR and lower carrier mobility in WO$_2$ are attributed to the coexistence of Dirac fermions and ordinary carriers. As listed in Table I, topological semimetals with smaller MR and lower carrier mobility tend to have larger carrier densities. Cd$_3$As$_2$ and NbP exhibit small carrier densities of the order of $10^{18}$ cm$^{-3}$ because only Dirac or Weyl fermions contribute to the transport properties. On the other hand, in $\beta$-ReO$_2$, ordinary carriers forming large Fermi surface coexist with Dirac fermions originating from Dirac loop chain [12]. Our *ab-initio* calculations revealed the coexistence of ordinary carriers and Dirac fermions also in WO$_2$. The carrier mobility estimated from the transport properties is an averaged value of all the carriers including normal carriers and high-mobility Dirac fermions. Thus, the obtained carrier mobility should be significantly lower than that of the Dirac fermions present.

The large MR of nonmagnetic metals and semimetals is usually attributed to classical carrier compensation [4] [34] [38]. In addition to this mechanism, the topological electronic structure has been discussed as a cause [1] [41] [42] [43] [44]. When linear dispersion is present near $E_F$, backscattering of the quasiparticles is strongly suppressed, and these quasiparticles contribute significantly to the transport properties as high-mobility carriers. When a magnetic field is applied, backscattering suppression is violated and the mobility of the carriers decreases drastically, resulting in a large increase in resistance.

Table I. Comparison of transport parameters and topological states among various materials exhibiting large MR.

| Material | Topological state | MR (%) | $\mu$ (cm$^2$ V$^{-1}$ s$^{-1}$) | $n$ (cm$^{-3}$) | Ref |
|---|---|---|---|---|---|
| Cd$_3$As$_2$ | Dirac semimetal | $1.3\times10^5$ (5 K, 9 T) | $8.7\times10^6$ (5 K) | $7.4\times10^{18}$ (5 K) | [1] |
| NbP | Weyl semimetal | $8.5\times10^5$ (1.85 K, 9 T) | $5.0\times10^6$ (1.85 K) | $1.5\times10^{18}$ (1.85 K) | [2] |
| ZrSiS | Nodal-line semimetal | $1.4\times10^5$ (2 K, 9 T) | $1.37\times10^4$ (2 K) | $3.6\times10^{20}$ (2 K) | [3] [45] |
| $\beta$-ReO$_2$ | Dirac loop chain metal | $2.2\times10^4$ (2 K, 10 T) | $3.8\times10^3$ (2 K) | $7.7\times10^{21}$ (2 K) | [12] |
| WO$_2$ | Nodal-line semimetal | $1.3\times10^4$ (2 K, 9 T) | $2.6\times10^4$ (5 K) | $2.3\times10^{20}$ (5 K) | This work |

Carrier compensation in WO$_2$ was revealed by Hall-effect measurements and electronic structure calculations. Electrons and holes coexist at low temperatures. The electron-to-hole density ratio, $n_h/n_e$, at 5 K was ~0.8. Compared with $n_h/n_e \sim 1$ in WTe$_2$ [4], where the compensation mechanism explains the large MR, WO$_2$ deviates from perfect compensation. When the electron and hole densities were unbalanced, the MR decreased rapidly [4]. Therefore, although the compensation mechanism should play some role in the large MR of WO$_2$, it is difficult to explain the large MR using only the compensation mechanism. Furthermore, in the case of the compensation mechanism, MR increases in proportion to the square of the magnetic field and saturates at high magnetic fields [34] [35]; the MR of WO$_2$ is proportional to $B^{1.6}$ and does not saturate until 9 T is reached, which differs from that expected for the compensation mechanism.

The signature of the topological electronic structure (Dirac points and DNLs) appeared in the transport properties at low temperatures. In WO$_2$, the hole mobility increased rapidly at temperatures below 50 K and reached a considerably high mobility of 25,700 cm$^2$ V$^{-1}$ s$^{-1}$ at 5 K. In other topological semimetals, the carrier mobility also increases rapidly at low temperatures, and a large MR is observed [38]. As discussed above, in WO$_2$, high-mobility carriers originating from the linear dispersion coexist with carriers originating from other ordinary bands. At high temperatures, carriers originating from ordinary bands with a large carrier density should have a dominant contribution to the transport properties, whereas at low temperatures, where the mobility of high-mobility carriers increases, they have a dominant contribution to the transport properties, despite their small carrier density. The deviation from the Kohler rule at < 50 K and the change in the MR anisotropy are attributed to the crossover of the dominant carriers from ordinary to high-mobility carriers. Similar changes in the dominant carrier to the transport properties have been noted in $\beta$-ReO$_2$ [12]. The large MR at low temperatures, where high-mobility carriers dominantly contribute to the transport properties, suggests that Dirac electrons play an important role in achieving a large MR. Our recent quantum oscillation study using magnetic torque measurements supports the coexistence of trivial carriers and Dirac fermions. The observed quantum oscillations originate from both ordinary carriers and carriers with significantly small cyclotron masses, which are likely to originate from linear dispersion.

In WO$_2$, major parts of the DNLs disappear in the presence of SOC, and gaps open at the band intersections. This band anticrossing induced by the SOC results in a large intrinsic SHE. In IrO$_2$, the anticrossing region of the DNL near $E_F$ was responsible for the large SHE [8]. In WO$_2$, the DNLs without SOC, which create anticrossing near $E_F$, may also contribute to the large SHE.

V. CONCLUSIONS

We grew high-quality single crystals of WO$_2$ with a distorted-rutile-type structure that exhibits a large MR of 13,200% at 2 K and 9 T. By fitting the Hall conductivity, we obtained $n_h = 2.33 \times 10^{20}$ cm$^{-3}$ and $n_e = 1.9 \times 10^{20}$ cm$^{-3}$ at 5 K. The hole mobility increased rapidly at low temperatures and reached a large value of 25,700 cm$^2$ V$^{-1}$ s$^{-1}$ at 5 K. These unique transport properties indicate the presence of Dirac electrons originating from the linear dispersion in WO$_2$. Band calculations showed that WO$_2$ had two types of dispersive nodal lines when the SOC was not considered and most of these lines disappeared when the SOC was considered. However, at high-symmetry points in the BZ and on the $k_y = \pi/b$ plane, Dirac points and DNLs near $E_F$ did not disappear even when SOC was considered. The transport properties of WO$_2$ are attributed to these Dirac electrons. WO$_2$ is a thermally and chemically stable oxide material, and the present results demonstrate promising transport properties originating from its topological electronic structure. The relationship between the large spin Hall effect and the topological electronic structure should be investigated in future work.

ACKNOWLEDGMENTS

This study was partially conducted under the Visiting Researcher's Program of the Institute for Solid State Physics, the University of Tokyo. The single-crystal XRD experiment was conducted at the BL02B1 of SPring-8, Hyogo, Japan (Proposal No. 2022B0607). We would like to thank Editage (www.editage.jp) for editing the manuscript. This study was partially supported by the Japan Society for the Promotion of

Science (KAKENHI Grant Nos. JP20H01858, JP23H04860, JP24K00758, and JP24H01187).DATA AVAILABILITY

The data that supports the findings of this study are available within the article and its supplementary material.

**REFERENCES**

[1] T. Liang, Q. Gibson, M. N. Ali, M. Liu, R. J. Cava, and N. P. Ong, *Ultrahigh Mobility and Giant Magnetoresistance in the Dirac Semimetal Cd3As2*, Nat Mater **14**, (2015).

[2] C. Shekhar et al., *Extremely Large Magnetoresistance and Ultrahigh Mobility in the Topological Weyl Semimetal Candidate NbP*, Nat Phys **11**, (2015).

[3] R. Singha, A. K. Pariari, B. Satpati, and P. Mandal, *Large Nonsaturating Magnetoresistance and Signature of Nondegenerate Dirac Nodes in ZrSiS*, Proc Natl Acad Sci U S A **114**, (2017).

[4] M. N. Ali et al., *Large, Non-Saturating Magnetoresistance in WTe2*, Nature **514**, (2014).

[5] X. Huang et al., *Observation of the Chiral-Anomaly-Induced Negative Magnetoresistance: In 3D Weyl Semimetal TaAs*, Phys Rev X **5**, (2015).

[6] J. N. Nelson et al., *Dirac Nodal Lines Protected against Spin-Orbit Interaction in IrO2*, Phys Rev Mater **3**, (2019).

[7] X. Xu et al., *Strong Spin-Orbit Coupling and Dirac Nodal Lines in the Three-Dimensional Electronic Structure of Metallic Rutile IrO2*, Phys Rev B **99**, (2019).

[8] Y. Sun, Y. Zhang, C. X. Liu, C. Felser, and B. Yan, *Dirac Nodal Lines and Induced Spin Hall Effect in Metallic Rutile Oxides*, Phys Rev B **95**, (2017).

[9] K. Fujiwara, Y. Fukuma, J. Matsuno, H. Idzuchi, Y. Niimi, Y. Otani, and H. Takagi, *5diridium Oxide as a Material for Spin-Current Detection*, Nat Commun **4**, (2013).

[10] K. Ueda, N. Moriuchi, K. Fukushima, T. Kida, M. Hagiwara, and J. Matsuno, *Spin-Orbit Torque Generation in $\mathrm{Ni}\mathrm{Fe}/\mathrm{Ir}{\mathrm{O}}_{2}$ Bilayers*, Phys Rev B **102**, (2020).

[11] S. S. Wang, Y. Liu, Z. M. Yu, X. L. Sheng, and S. A. Yang, *Hourglass Dirac Chain Metal in Rhenium Dioxide*, Nat Commun **8**, (2017).

[12] D. Hirai, T. Anbai, S. Uji, T. Oguchi, and Z. Hiroi, *Extremely Large Magnetoresistance in the Hourglass Dirac Loop Chain Metal β-ReO2*, J Physical Soc Japan **90**, (2021).

[13] D. Hirai et al., *Fermi Surface and Light Quasi Particles in Hourglass Nodal Chain Metal β-ReO2*, Journal of Physics Condensed Matter **35**, (2023).

[14] D. B. Rogers, R. D. Shannon, A. W. Sleight, and J. L. Gillson, *Crystal Chemistry of Metal Dioxides with Rutile-Related Structures*, Inorg Chem **8**, (1969).

[15] F. H. Jones, R. G. Egdell, A. Brown, and F. R. Wondre, *Surface Structure and Spectroscopy of WO2(012)*, Surf Sci **374**, (1997).

[16] Z. Hiroi, *Structural Instability of the Rutile Compounds and Its Relevance to the Metal-Insulator Transition of VO2*, Progress in Solid State Chemistry.

[17] Q. Chen et al., *Large Magnetoresistance and Nonzero Berry Phase in the Nodal-Line Semimetal Mo O2*, Phys Rev B **102**, (2020).

[18] K. Ueda, H. Fujii, T. Kida, M. Hagiwara, and J. Matsuno, *Spin Current Generation from an Epitaxial Tungsten Dioxide WO2*, APL Mater **11**, (2023).

[19] T. Takayama, J. Chaloupka, A. Smerald, G. Khaliullin, and H. Takagi, *Spin⇓orbit-Entangled Electronic Phases in 4d and 5d Transition-Metal Compounds*, Journal of the Physical Society of Japan.

[20] G. Kresse and J. Hafner, *Ab Initio Molecular Dynamics for Liquid Metals*, Phys Rev B **47**, (1993).

[21] G. Kresse and J. Furthmüller, *Efficient Iterative Schemes for Ab Initio Total-Energy Calculations Using a Plane-Wave Basis Set*, Phys Rev B Condens Matter Mater Phys **54**, (1996).

[22] J. P. Perdew, K. Burke, and M. Ernzerhof, *Generalized Gradient Approximation Made Simple*, Phys Rev Lett **77**, (1996).

[23] H. J. Monkhorst and J. D. Pack, *Special Points for Brillouin-Zone Integrations*, Phys Rev B **13**, (1976).

[24] G. Pizzi et al., *Wannier90 as a Community Code: New Features and Applications*, Journal of Physics Condensed Matter **32**, (2020).

[25] M. Kawamura, *FermiSurfer: Fermi-Surface Viewer Providing Multiple Representation Schemes*, Comput Phys Commun **239**, (2019).

[26] R. Hartman, *Temperature Dependence of the Low-Field Galvanomagnetic Coefficients of Bismuth*, Physical Review **181**, (1969).

[27] F. F. Tafti, Q. D. Gibson, S. K. Kushwaha, N. Haldolaarachchige, and R. J. Cava, *Resistivity Plateau and Extreme Magnetoresistance in LaSb*, Nat Phys **12**, (2016).

[28] N. Kumar et al., *Extremely High Magnetoresistance and Conductivity in the Type-II Weyl Semimetals WP2 and MoP2*, Nat Commun **8**, (2017).

[29] L. K. Zeng et al., *Compensated Semimetal LaSb with Unsaturated Magnetoresistance*, Phys Rev Lett **117**, (2016).

[30] J. He et al., *Distinct Electronic Structure for the Extreme Magnetoresistance in YSb*, Phys Rev Lett **117**, (2016).
7